\documentclass[]{interact}

\usepackage{epstopdf}
\usepackage[caption=false]{subfig}

\usepackage[numbers,sort&compress]{natbib}
\bibpunct[, ]{[}{]}{,}{n}{,}{,}
\makeatletter
\def\NAT@def@citea{\def\@citea{\NAT@separator}}
\makeatother

\theoremstyle{plain}

\theoremstyle{definition}

\theoremstyle{remark}

\usepackage{color}

\begin{document}

\articletype{ARTICLE TEMPLATE}

\title{Hydrogen bond analysis of confined water in mesoporous silica using the reactive force field}

\author{
\name{Tomoko Mizuguchi\textsuperscript{a}\thanks{CONTACT T. Mizuguchi. Email: mizuguti@kit.ac.jp} and Katsumi Hagita\textsuperscript{b} and Susumu Fujiwara\textsuperscript{a} and Takeshi Yamada\textsuperscript{c}}
\affil{\textsuperscript{a}Faculty of Materials Science and Engineering, Kyoto Institute of Technology, Matsugasaki, Sakyo-ku, Kyoto 606-8585, Japan; \textsuperscript{b}Department of Applied Physics, National Defense Academy, 1-10-20, Hashirimizu, Yokosuka, Kanagawa 239-8686, Japan; \textsuperscript{c}CROSS Neutron Science and Technology Center, IQBRC Bldg, 162-1 Shirakata, Tokai, Naka, Ibaraki 319-1106, Japan 
}
}

\maketitle

\begin{abstract}
The structural and dynamical properties of water confined in nanoporous silica with a pore diameter of 2.7 nm were investigated by performing large-scale molecular dynamics simulations using the reactive force field.
The radial distribution function and diffusion coefficient of water were calculated, and the values at the center of the pore agreed well with experimental values for real water.
In addition, the pore was divided into thin coaxial layers, and the average number of hydrogen bonds, hydrogen bond lifetime, and hydrogen bond strength were calculated as a function of the radial distance from the pore central axis.
The analysis showed that hydrogen bonds involving silanol (Si-OH) have a longer lifetime, although the average number of hydrogen bonds per atom does not change from that at the pore center.
The longer lifetime, as well as smaller diffusion coefficient, of these hydrogen bonds is attributed to their greater strength.
\end{abstract}


\section{Introduction}

Understanding the structural and dynamical properties of water in confinement is essential to many scientific fields and technological applications, such as permeation in the ion channels of a biological membrane, capacitance of an electrical double layer in fuel cells, and controlled drug release~\cite{Murata2000,Sui2001,Chmiola2006,Cavallaro2004}.
Mesoporous materials have attracted attention in recent years because their high specific surface area, large specific pore volume, and narrow pore size distribution allow a wide range of possible industrial applications.
In particular, mesoporous silicate MCM-41, which has an amorphous structure, is one of the most widely used mesoporous material~\cite{Bhattacharyya2006} and has potential application in
drug delivery~\cite{Vallet-Regi2001,Colilla2013,Trewyn2007}, in which water plays an important role~\cite{Cavallaro2004}.
In confined systems, surface properties have a significant impact on the physical properties of the confined substance.
However, water behavior near an amorphous surface is much more difficult to investigate than that near a crystalline surface because the topology of an amorphous surface is ill-defined at the molecular level.

The behavior of confined water has been widely studied by both theoretical~\cite{Renou2014,Giovambattista2009,Lerbret2011,Chen2014,Alexiadis2008,Gallo2000,Rovere1998,Collin2018,Bourg2012,Cicero2008,Corry2008,Takaiwa2008,Bai2006,Neek-Amal2016,Koga2001,Huang2006,Zhu2013} and experimental~\cite{Aso2013,Mancinelli2009,Chu2007,Faraone2004,Kyakuno2011,Yamada2011} techniques in the past few decades.
These works have indicated that the properties of confined water may strongly differ from that of bulk water depending on surface properties and size of confinement space.
It is thus important to properly evaluate the influence of confinement on water.

In recent years, the dynamical structure factor of water in mesoporous silica has been measured by quasi-elastic neutron scattering (QENS), and the self-diffusion coefficients ($D$) of water in fast and slow modes were evaluated by dividing the space into two regions. 
Yamada {\it et al.} reported that $D$ significantly differs between the vicinity of the interface and pore center~\cite{privatecom}. 
However, the method of dividing the space is empirical, and its relationship with the microscopic atomic structure and dynamics is not sufficiently confirmed. 
This question is expected to be resolved through the analysis of trajectories from molecular dynamics (MD) simulations. 

In MD simulations, the choice of force field is very important in reasonably estimating the interfacial effect. 
Recently, Bourg and Steefel reproduced the experimental diffusion constant obtained by QENS through MD simulations using the Clay force field (CLAYFF)  and extended simple point charge (SPC/E) water model~\cite{Bourg2012}. 
They noted that the use of an {\it ab initio} method ~\cite{Musso2012,Cimas2014,Lowe2015} or the reactive force field (ReaxFF) in MD simulations allows a good description of proton transfer reactions.
van Duin and co-workers developed the ReaxFF package~\cite{Vanduin2001}, which has been applied in various research areas~\cite{Senftle2016}. 
In recent years, MD simulations of the silica/water interface have been performed using ReaxFF~\cite{Fogarty2010,Leroch2012,Yeon2016,Rimsza2016,Rimsza2017}. 
This method has a lower computational cost than quantum-mechanics-level calculations such as {\it ab initio} MD simulations. 
A drawback of MD simulations using classical force fields is that an appropriate chemical structure must be created before starting the simulation, and this does not change during the simulation because the chemical bond is fixed.
This problem is particularly conspicuous when dealing with interfacial systems.
Using ReaxFF, the physical properties of confined water can be obtained with high accuracy while avoiding the problem of interaction between water and silica.

In this study, we performed MD simulations of water in mesoporous silica using ReaxFF and analyzed water dynamics to clarify the influence of the surface on the physical properties of water at the atomic and molecular levels. Specifically, the hydrogen bond (H-bond) dynamics in the inhomogeneous system composed of amorphous silica (a-SiO$_2$) and water was predicted.
To obtain sufficient statistics, we used a large system since the pore was analyzed by dividing it into thin layers.

The remainder of this paper is organized as follows. 
The next section provides an overview of ReaxFF potentials and details of the preparation of the silica-water system. 
Section~\ref{sec:result} presents the structural properties and diffusion of water in nanoporous silica and discusses the results of the analyses of H-bond configurations and dynamics. 
The final section summarizes the main findings and consequent conclusions.

\section{\label{sec:method}Methods}
\subsection{\label{sec:reaxff}ReaxFF potentials}

To account for bond breakage and formation, we performed MD simulations using the ReaxFF potential~\cite{Vanduin2001}.
In ReaxFF, all bond orders are calculated directly from interatomic distances every time step.
The calculation of bond-order-based terms in the potential allows for bond breakage and formation.
Note that H-bonds are explicitly defined in the ReaxFF potential.
Additionally, the charge of each atom is dynamically assigned by a charge equilibration method~\cite{Rappe1991}.
Detailed explanation of ReaxFF can be found in previous articles~\cite{Vanduin2001,Chenoweth2008,Aktulga2012,Senftle2016}.

We used the ReaxFF parameter set for silica-water system developed by Yeon and van Duin in 2015 to simulate hydrolysis reactions at the SiO$_2$/water interface~\cite{Yeon2016}.
It has been used to study interactions between water and nanoporous silica and has yielded results that agreed well with {\it ab initio} MD data for reaction mechanisms, hydroxylation rates, defect concentrations, and activation energies~\cite{Rimsza2016}.
Thus, we chose this parameter set to study water dynamics in mesoporous silica.
However, this parameter set was also reported to yield an unrealistically high fraction of pentacoordinated Si atoms in glassy silica~\cite{Yu2018}.
Yu {\it et al.} reported that both reactivity and hydrophilicity are influenced by the atomic topology of the surface of the glassy silica~\cite{Yu2018}.
Therefore, we also used the ReaxFF parameter set developed by Pitman {\it et al.}~\cite{Pitman2012,Fogarty2010}, which provided an atomic structure of glassy silica that agreed well with the experimental structure~\cite{Yu2016,Yu2017}.
ReaxFF also allows O--H bond extension/contraction and changes in the H--O--H angle of water during MD simulations, unlike traditional force fields for water that treat the water molecule as rigid.

\subsection{\label{sec:structure}Preparation of the silica-water system}
The system composed of water in nanoporous silica was created as follows. 
First, bulk a-SiO$_2$ was prepared at 3000 K and 1 atm in the {\it NP$\rm_z$T} ensemble (i.e., constant number of atoms, pressure along the {\it z}-direction with fixed {\it xy}-area, and temperature) using the Morse potential~\cite{Demiralp1999}. The {\it z}-direction is parallel to the axis of a cylindrical pore that is subsequently created. 
After switching to ReaxFF MD simulation, the system was relaxed for 200 ps at 3000 K and 1 atm in the {\it NP$\rm_z$T} ensemble, then quenched to 300 K at a cooling rate of 15 K/ps in the {\it NVT} ensemble.  
After a 300-ps relaxation at 300 K and 1 atm in the {\it NPT} ensemble, 4 cylindrical pores with a diameter of 2.7 nm were created by removing atoms. Furthermore, a few oxygen atoms at the pore surface were removed to obtain a Si/O ratio of 1:2. 
Water was then placed inside the pore, and a 20-ps MD simulation was performed only for water using the TIP3P (transferable intermolecular potential with 3 points) model~\cite{Jorgensen1983} while silica was immobilized.
The system consists of an 8.9 $\times$ 7.7 $\times$ 45.1 nm rectangular box with periodic boundary conditions employed in all directions.
Finally, ReaxFF MD simulation of the silica-water system was conducted.
Silanols (Si-OH) rapidly formed at the interface through chemical reactions between silica and water.
After a 500-ps equilibration run at 300 K and 1 atm, a 50-ps production run was performed for analyses.
All simulations were conducted using the Nos{\'e}-Hoover thermostat and barostat using the Large-scale Atomic/Molecular Massively Parallel Simulator (LAMMPS) software~\cite{Plimpton1995}. The ReaxFF MD simulations were performed using a time step of 0.25 fs through the USER-REAXC package of LAMMPS.

\section{\label{sec:result}Results and Discussions}
\subsection{Structural properties}
\begin{figure}
\includegraphics[width=14cm]{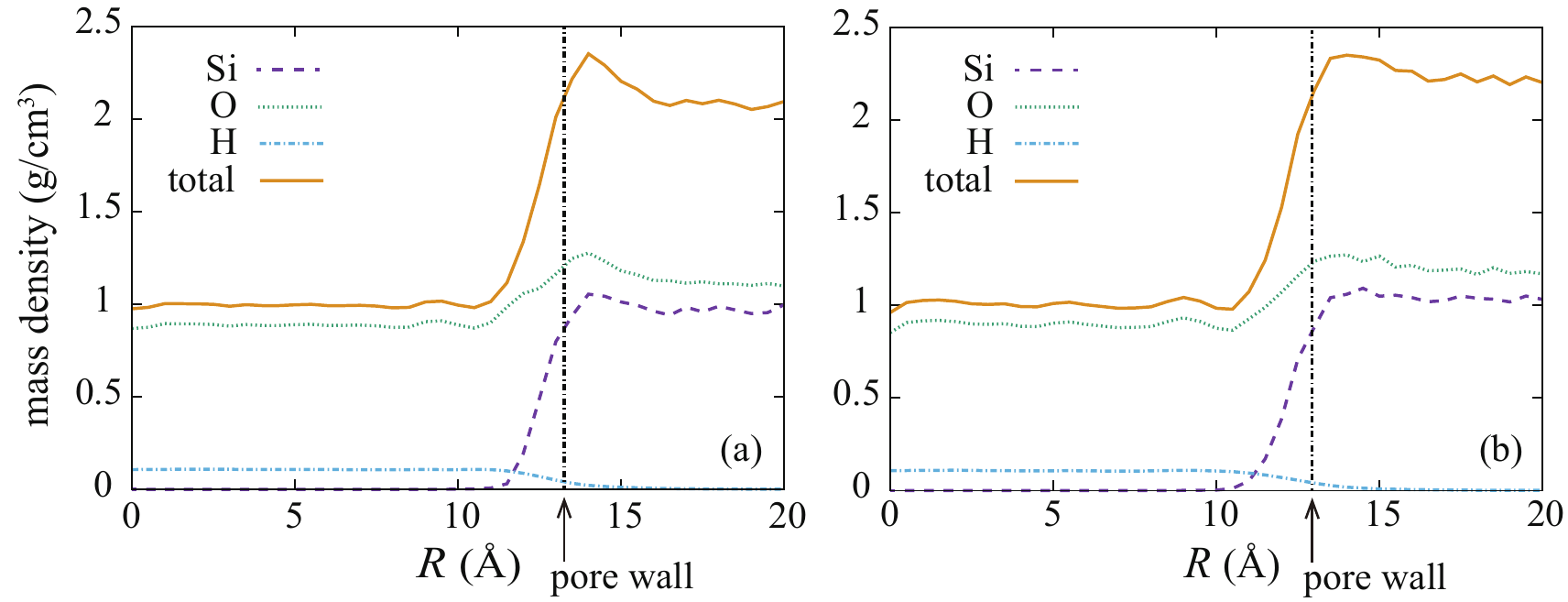}
\caption{Mass density profiles of silica-water system at 300 K calculated using (a) Yeon's and (b) Pitman's parameter sets [oxygen (green dotted line), hydrogen (blue dash-dotted line), silicon (purple dashed line), total (orange solid line)]. The location of the pore wall ($R_{\rm Si-in}$) is also displayed as a black dash-dotted line.}
\label{f2}
\end{figure}

Figure~\ref{f2} shows the average mass density of the silica-water system at 300 K, derived from MD simulations using Yeon's~\cite{Yeon2016} and Pitman's~\cite{Pitman2012} ReaxFF parameter sets, as a function of the radial distance from the pore central axis ({\it R}).
It was calculated by dividing the cylindrical pore space into hollow cylinders with a 0.5-\AA\ thickness, determining the density of each layer, and taking the time average over the 20-ps simulation. 
The approximate pore radius is also shown in Fig.~\ref{f2}.  
It was calculated by slicing the cylindrical pore space along the $z$-axis into thin disks with a 0.2-\AA\ thickness, determining the $R$ of the innermost  Si atom ($R_{\rm Si-in}$) in each disk, and taking the average of all values over the 20-ps time period. 
The $R_{\rm Si-in}$ values, calculated using Yeon's and Pitman's parameter sets, are 13.24 and 12.85 \AA, respectively.
The density distribution of Si, which represents the roughness of the amorphous surface, extends to $R=$ 11 (Yeon) and 10 \AA\ (Pitman).
Muroyama {\it et al} measured the surface roughness of MCM-41 by using powder X-ray diffraction and concluded that the surface roughness on the silica wall is 2 \AA\ for the as-synthesized hexagonal pore and 1 \AA\ for the calcined circular pore~\cite{Muroyama2006}.
They defined the surface roughness as the distance from the boundary between the pores and the wall, which is determined using a threshold density ratio within the pore to that of the wall,  to a point of a half value of the boundary density between the pores and the wall.
We determined the surface roughness according to their definition.
The boundary in this work is defined as the point where the density of silicon reach the bulk value and is located at $R=$13.8 \AA\ and 13.6 \AA\ for Yeon's and Pitman's parameter sets.
The points of a half value of the boundary density are $R=$12.7 \AA\ and 12.4 \AA\ respectively, and thus the surface roughnesses are 1.1 \AA\ and 1.2 \AA\ for Yeon's and Pitman's parameter sets.
These values are coincident with the experimental value of 1 \AA\ for the calcined circular pore.

The densities of bulk a-SiO$_2$ (averaged over $R=$ 17--20 \AA), calculated using Yeon's and Pitman's parameter sets, are 2.08 and 2.21 g/cm$^3$, respectively.
These values coincide with those obtained in previous studies: 2.10~\cite{Rimsza2017} and 2.18~\cite{Yu2018} g/cm$^3$ at 300 K using Yeon's and Pitman's parameter sets, respectively.
The density of a-SiO$_2$ calculated using Pitman's parameter set is higher than that calculated using Yeon's parameter set and closer to the experimental value of 2.2 g/cm$^3$~\cite{Mozzi1969}.
Given the same number of water molecules confined in the pores, the two models yield different densities for water; the density in Yeon's model is lower than that in Pitman's model.
Thus, in Yeon's and Pitman's models, about 8364 and 7724 water molecules are confined in each pore for a total of 33457 and 30896 water molecules (since there are four pores in each system), respectively, to make the water densities at the pore center almost equal.
The average densities of water at $R <$5 \AA\ are 0.993 and 1.006 g/cm$^3$ in Yeon's and Pitman's models, respectively.
Note that the density of water at 300 K and 1 atm is 0.997 g/cm$^3$.

\begin{figure}
\includegraphics[width=14cm]{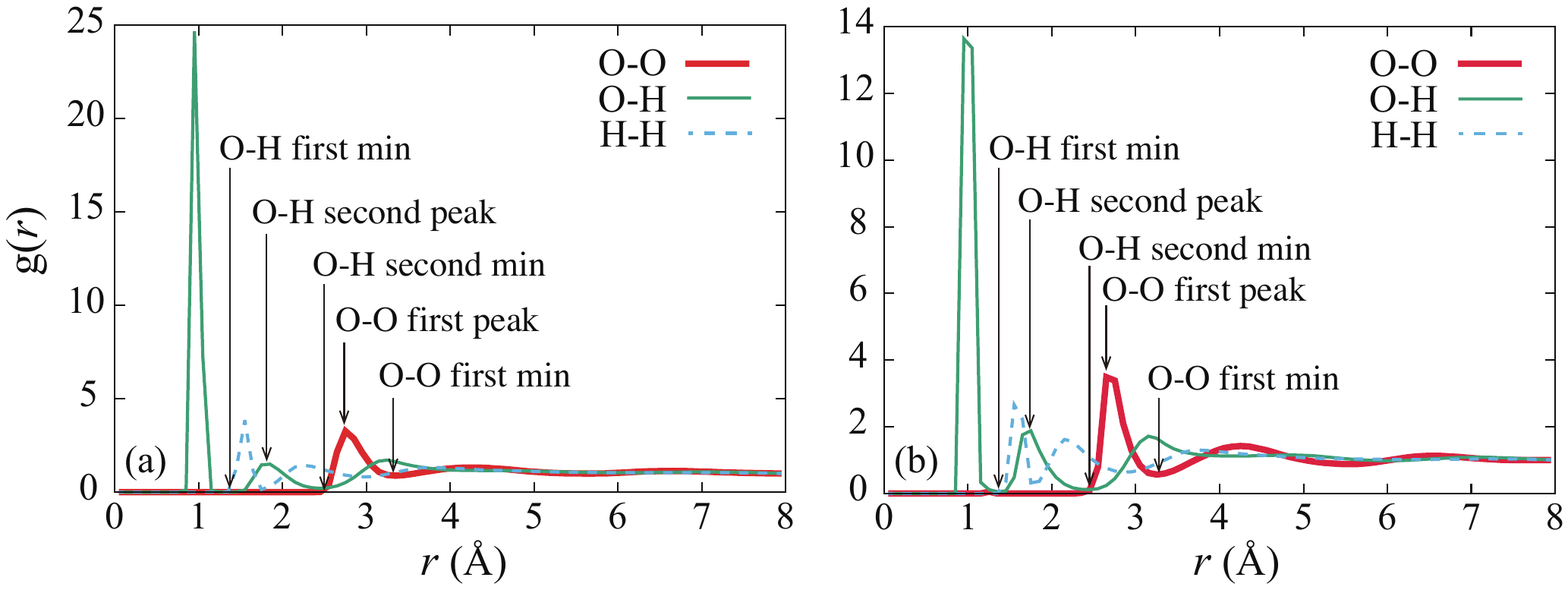}
\caption{O--O (red bold solid line), H--H (blue dashed line), and O--H (green solid line) radial distribution functions at 300 K calculated using (a) Yeon's and (b) Pitman's parameter sets. Calculations were performed for atoms located at $R <$12 \AA.}
\label{f3}
\end{figure}

The radial distribution function (RDF) between two atoms is defined as
\begin{equation}\label{eq1}
g_{12}(r) = \frac{\langle n_2(r) \rangle}{4\pi r^2 \rho_2 \Delta r},
\end{equation}
where $r$ is the distance between a pair of type 1 and 2 atoms, $n_2(r)$ is the number of type 2 atoms in the shell between $r$ and $r+\Delta r$ around type 1 atoms and $\rho_2$ is the density of type 2 atoms.
For example, in the O--H RDF, the type 1 and 2 atoms are O and H, respectively.
The term in angular brackets represents the average over all atoms of interest and different starting times of the RDF calculation.
To calculate the RDFs of water in the nanopore, atoms located at $R <$4 \AA\ were taken as the type 1 atoms in Eq.~\ref{eq1}, and the cut-off radius for the $g(r)$ calculation was 8.0 \AA.
Therefore, the RDF calculations were performed for atoms located at $R <$12 \AA, excluding those near the interface.
The RDFs calculated using Yeon's and Pitman's parameter sets are displayed in Fig.~\ref{f3}.
Note that the first peak of the O--H RDF is the intramolecular O--H distance, while the second one represents the nearest intermolecular O--H distance. 
The locations of the peaks and minima in RDFs are listed in Table~\ref{tb1} with experimental values for comparison.
The results in bulk water using classical water models are also listed as reference.
The RDFs calculated using Yeon's parameter set agree well with experimental values, while the water structure obtained using Pitman's parameter set deviates from that of real water.
The locations of the first minimum in the O--O and O--H RDFs and second minimum in the O--H RDF were used for H-bond analysis, as described in Section~\ref{sec:hb}.

\begin{table}
\tbl{Locations of the peaks and minima in the radial distribution functions (RDFs) of water atoms located at $R <$12 \AA\ at 300 K. Data in experiment and with classical water models are listed for comparison.\label{tb1}}
{\begin{tabular}{lcccccc}
RDF (\AA) & Yeon & Pitman & Experimental\textsuperscript{a} & TIP3P-Ew\textsuperscript{b} & TIP4P/2005\textsuperscript{c} & SPC/E\textsuperscript{d}\\ \hline
O--O first peak & 2.76 & 2.69 & 2.79 & 2.71 & 2.79 & 2.73\\
O--O first min. & 3.35 & 3.24 & 3.39 & 3.31 & 3.37 & 3.29\\
O--H second peak\textsuperscript{e} & 1.81 & 1.70 & 1.86 & 1.79 & 1.85 & 1.73\\
O--H second min. & 2.45 & 2.45 & 2.46 & 2.37 & 2.45 & 2.39\\ 
\end{tabular}}
\tabnote{\textsuperscript{a}Experimental data at 298 K derived from Ref.~\cite{Soper2013}.\textsuperscript{b}Ref.~\cite{Price2004}.\textsuperscript{c}Ref.~\cite{Abascal2005}.\textsuperscript{d}Ref.~\cite{Berendsen1987}.\textsuperscript{e}It shows the nearest intermolecular O-H distance.}
\end{table}

\subsection{Water diffusion in nanoporous silica}
\begin{figure}
\includegraphics[width=14cm]{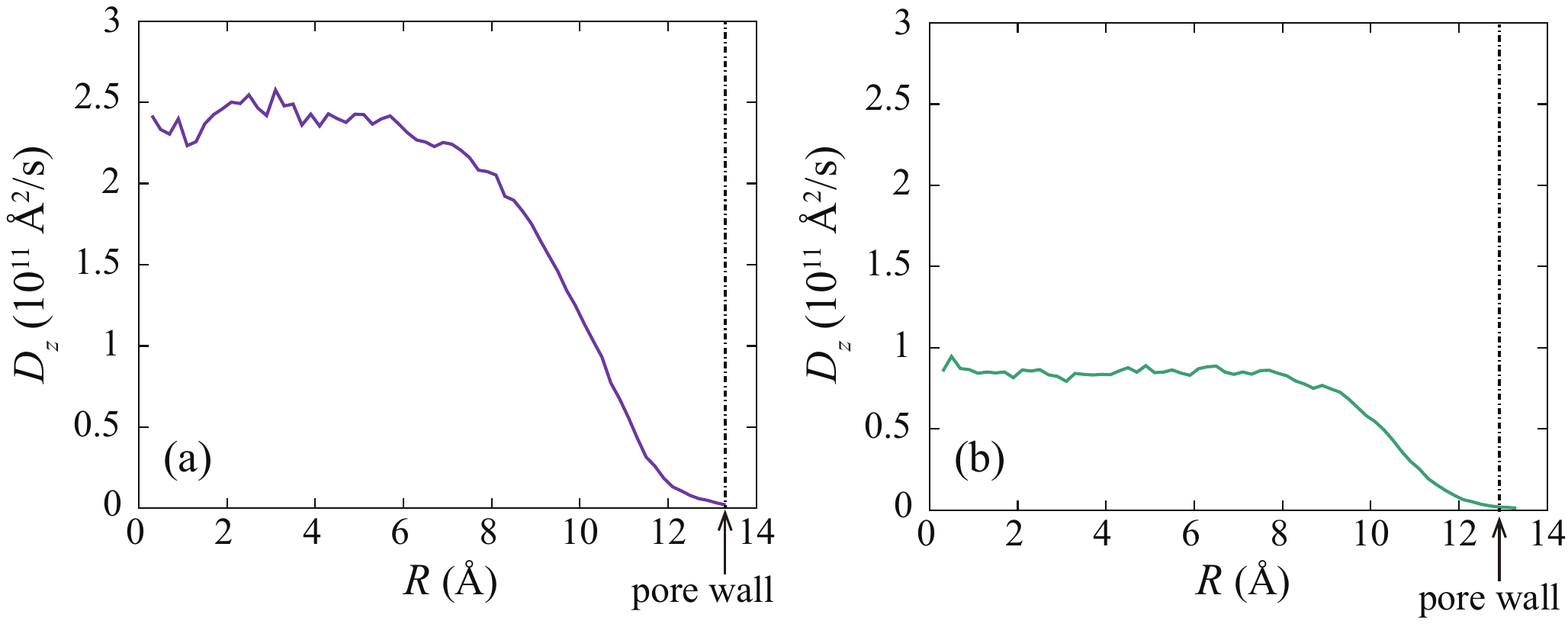}
\caption{Self-diffusion coefficient along the {\it z}-axis ($D\rm_z$) of oxygen atoms in the pores as a function of radial distance from the pore central axis ($R$). (a) Yeon's and (b) Pitman's parameter sets. The location of the pore wall ($R_{\rm Si-in}$) is also displayed as a black dash-dotted line.}
\label{f4}
\end{figure}

The {\it D} of atoms can be calculated from the long-time behavior of the mean square displacement (MSD) using the Einstein relation:
\begin{equation}\label{eq2}
D = \lim_{t \to \infty} \frac{\langle | \vec{r}(t) - \vec{r}(0) |^2 \rangle}{2dt},
\end{equation}
where {\it d} is the dimension and the term in angular brackets represents the average over all atoms of interest and different starting times of the $D$ calculation.
Applying this equation to our system, however, is nontrivial because $D$ varies with the distance from the interface.
On the basis of previous studies on water confined in nanospace~\cite{Kerisit2009,Bourg2012}, we divided the pore space into hollow cylinders with a 0.2-\AA\ thickness and calculated the {\it D} in each layer using the slope of MSD from {\it t} = 5 ps to {\it t} = 10 ps in the {\it z}-direction.
For this layer thickness, this time scale was found to be sufficiently long for the molecular motion of water to be in the diffusive regime, but sufficiently short for individual water molecules not to probe regions with very different {\it D}~\cite{Bourg2012}.
The MSDs in the {\it z}-direction were calculated for oxygen atoms in each region over the 20-ps time period.
The {\it D} along the {\it z}-direction ($D\rm_z$) was calculated from the MSD data using Eq.~\ref{eq2}with $d=$1 and is shown in Fig.~\ref{f4} as a function of {\it R}.
The {\it D} of bulk water is 2.3 $\times$ 10$^{11}$ \AA$^2$/s at 298 K based on several experiments~\cite{Holz2000}.
Bourg and Steefel evaluated the {\it D} of real water at 300 K as 2.41 $\times$ 10$^{11}$ \AA$^2$/s by interpolation from the {\it D} at 298 and 309 K using the Arrhenius relation~\cite{Bourg2012}.
Using Yeon's parameter set, the $D\rm_z$ averaged over values at $R <$6 \AA\ is 2.41 $\times$ 10$^{11}$ \AA$^2$/s, which agrees with the experimental value for bulk water.
On the other hand, using Pitman's parameter set, the $D\rm_z$ is about 0.85 $\times$ 10$^{11}$ \AA$^2$/s in the inner area of the pore.
This value is much lower than that of bulk water, but corresponds well with the {\it D} of water in the clay-zeolite composite~\cite{Pitman2012} used to develop the parameter set.

Yamada {\it et al.} examined the water dynamics in mesoporous silica with a 2.7-nm pore using QENS~\cite{privatecom}.
The water dynamics was divided into two modes, slow and fast, and the {\it D} of the latter corresponds to that of bulk water.
Because the fast-mode water is presumably located near the pore center, water in the center of the 2.7-nm pore of mesoporous silica behaves like bulk water.
This agreement with the results obtained using Yeon's parameter set indicates that the parameter set is suitable for the investigation of the dynamics of water confined in mesoporous silica.
In addition, the RDFs obtained using Yeon's parameter set agrees better with the experimental values than those obtained using Pitman's parameter set.
Thus, the following sections only discuss the results calculated using Yeon's parameter set.
Note that Yu {\it{et al.}} demonstrated that both reactivity and hydrophilicity are determined by the atomic topology of the surface and Pitman's parameter set provides a better description of the glassy silica structure~\cite{Yu2018}.
Thus, it would be interesting to perform simulations using Pitman's parameter set for SiO$_2$ and other ReaxFF parameters for water.

\subsection{Hydrogen-bond configurations and dynamics}\label{sec:hb}
The locations of the first minimum in the O--O RDF and second minimum in the O--H RDF were generally used for H-bond analysis.
Note that the first and second peaks in the O--H RDF in Fig.~\ref{f3} represent the nearest intramolecular and intermolecular O--H distances, respectively.
The location of the first minimum in the O--H RDF was additionally used as a criterion for H-bonding since ReaxFF is composed of atom-based potentials.
Thus, the H-bond O$_1\cdots$H$_2$--O$_2$ is defined using the following criteria: 
\begin{eqnarray}
r_{\mathrm{O_1O_2}} \leq 3.35\ {\mathrm \AA}, \nonumber \\
1.35\ {\mathrm \AA}\leq r_{\mathrm{O_1H_2}} \leq 2.45\ {\mathrm \AA}, \\
r_{\mathrm{O_2H_2}} \leq 1.35\ {\mathrm \AA}, \nonumber \\
\angle {\mathrm{H_2O_2O_1}} \leq 30^\circ, \nonumber
\end{eqnarray}
where O$_1$ is the H-bond acceptor and H$_2$--O$_2$ is the H-bond donor.
The angular cut-off is widely accepted in the literature~\cite{Tamai1996,Guardia2005,Alexiadis2008,Antipova2013,Liu2018}.

\begin{figure}
\includegraphics[width=7.5cm]{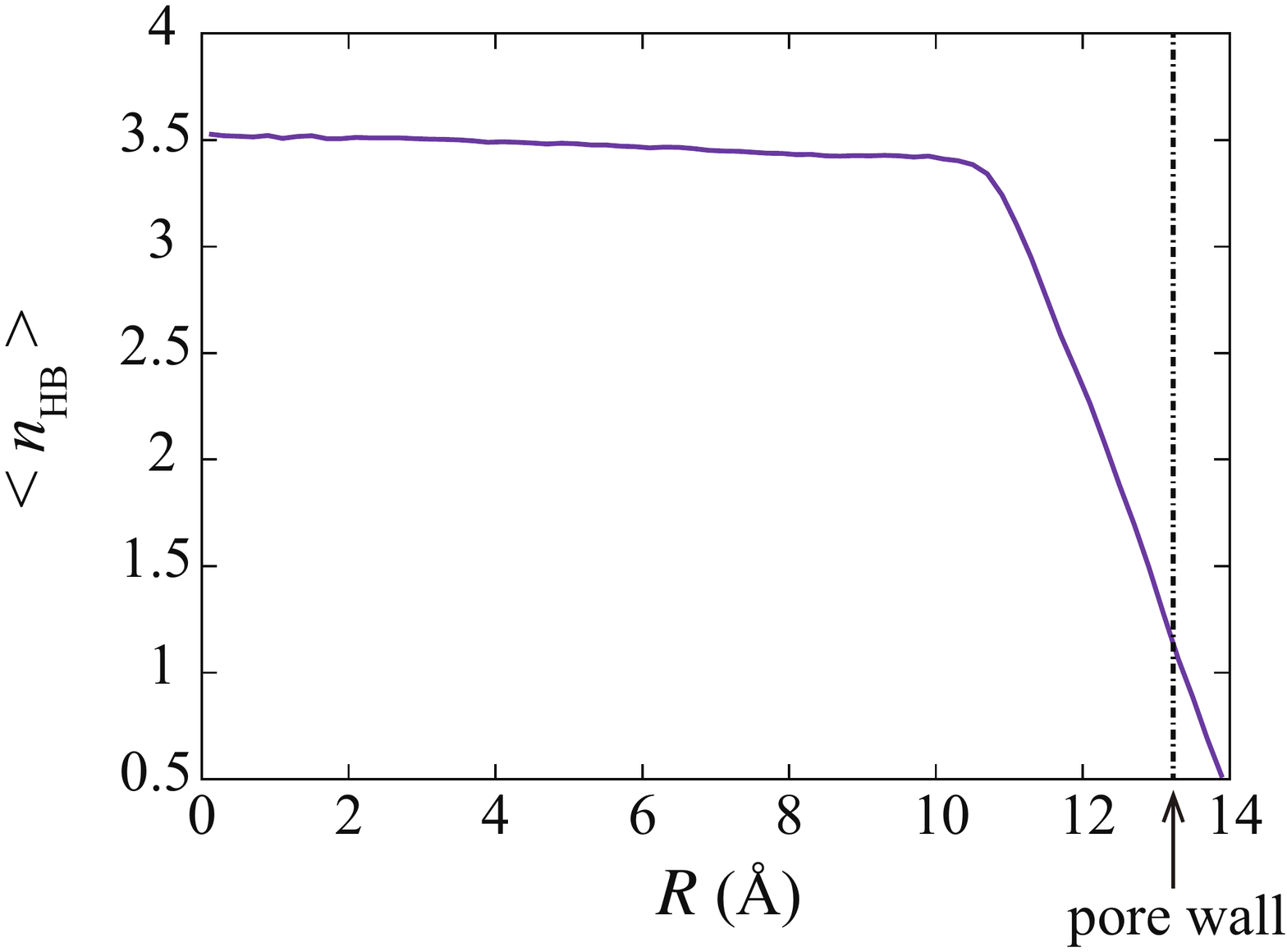}
\caption{Average number of hydrogen bonds per oxygen atom ($\langle n_{\rm HB} \rangle$) as a function of radial distance from the pore central axis ({\it R}). The location of the pore wall ($R_{\rm Si-in}$) is also displayed as a black dash-dotted line.}
\label{f5}
\end{figure}

First, the average number of H-bonds per oxygen atom ($\langle n_{\rm HB} \rangle$) was calculated and is shown in Fig.~\ref{f5} as a function of {\it R}.
The calculations were performed by dividing the pore space into hollow cylinders with a 0.2-\AA\ thickness.
In bulk water, the average number of H-bonds per water molecule is between 2 and 4 depending on the experimental technique~\cite{Rastogi2011}.
$\langle n_{\rm HB} \rangle$ in the center of the pore is about 3.5 at 300 K, which is a reasonable value for bulk-like water.
It is almost constant at $R <$11 \AA\ and decreases at $R >$11 \AA.
It indicates that the H-bond network becomes less structured at $R >$11 \AA.

The H-bond lifetime can be evaluated through the following autocorrelation function~\cite{Rapaport1983}
\begin{equation}\label{eq4}
C_x(t)=\left \langle \frac{\sum_{ij}s_{ij}(t+t_0)s_{ij}(t_0)}{\sum_{ij}s_{ij}^2(t_0)}\right \rangle_{t_0},
\end{equation}
where the subscript $x$ represents the two different definitions of lifetime: intermittent (I) and continuous (C).
In an intermittent lifetime, $s_{ij}$ is equal to 1 if a pair of atoms, {\it i} and {\it j}, are bonded at time {\it t} and 0 if the bond is absent.
Note that only pairs of atoms having $s_{ij}$ = 1 at {\it t} = 0 were included in the calculation.
In a continuous lifetime, $s_{ij}$ is allowed a single transition from 1 to 0 when the bond is first broken, but is not allowed to return to 1 if the same bond is reformed.
Namely, $C_{\rm I}$({\it t}) and $C_{\rm C}$({\it t}) focus on the elapsed time until the final and first breakage of the bond, respectively.
The term in angular brackets refers to the average over different starting times $t_0$ during the production run.
The calculations were performed by dividing the pore space into hollow cylinders with a 1.0-\AA\- thickness and only for acceptor oxygens remaining in each layer during the simulation.

The average H-bond lifetime $\tau_{\rm HB}$ can be evaluated by integrating $C_x$({\it t}) using the following equation~\cite{Antipova2013,Gowers2015},
\begin{equation}\label{eq5}
\tau_x=\int_0^\infty C_x(t) dt.
\end{equation}
To obtain the H-bond lifetime, the $C_x$({\it t}) calculated directly from the MD trajectories was fitted to the sum of three exponential functions and integrated analytically.

\begin{equation}\label{eq6}
C_x(t)=A_1{\mathrm{exp}}\left( -\frac{t}{\tau_1}\right)+A_2{\mathrm{exp}}\left(-\frac{t}{\tau_2}\right)+(1-A_1-A_2){\mathrm{exp}}\left(-\frac{t}{\tau_3}\right).
\end{equation}

\begin{figure}
\includegraphics[width=15cm]{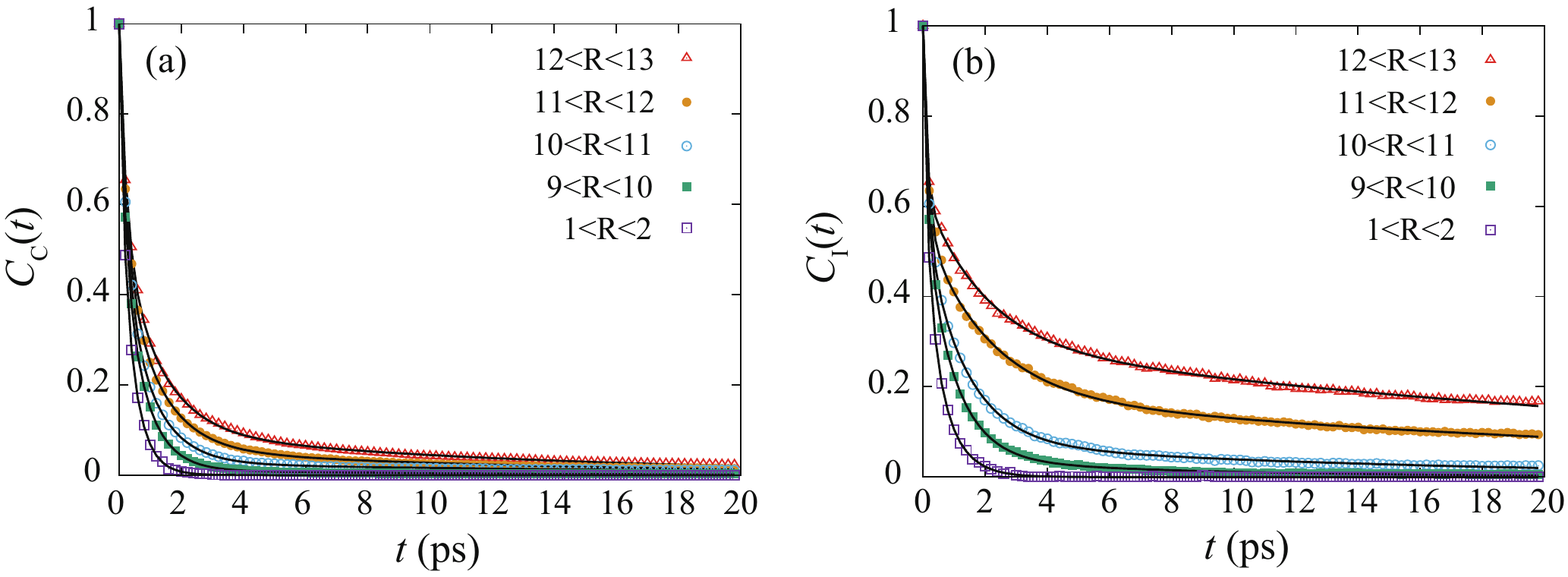}
\caption{(a) Continuous ($C_{\rm{C}}$({\it t})) and (b) intermittent ($C_{\rm{I}}$({\it t})) autocorrelation functions of H-bonds for several layers. Solid lines represent the fitting functions described in Eq.~\ref{eq6}.}
\label{f6}
\end{figure}

The calculated $C_x$({\it t}) for several layers are shown in Fig.~\ref{f6} with the fitting functions.
For the center of the pore, both $C_{\rm{I}}$({\it t}) and $C_{\rm{C}}$({\it t}) can be fitted to the sum of two exponential functions.
However, for the interface, the sum of three exponential functions show better fitting than the sum of two exponentials to both $C_{\rm I}$({\it t}) and $C_{\rm C}$({\it t}).
Thus, $C_x$({\it t}) was fitted to three exponentials for all layers.

\begin{figure}
\includegraphics[width=8cm]{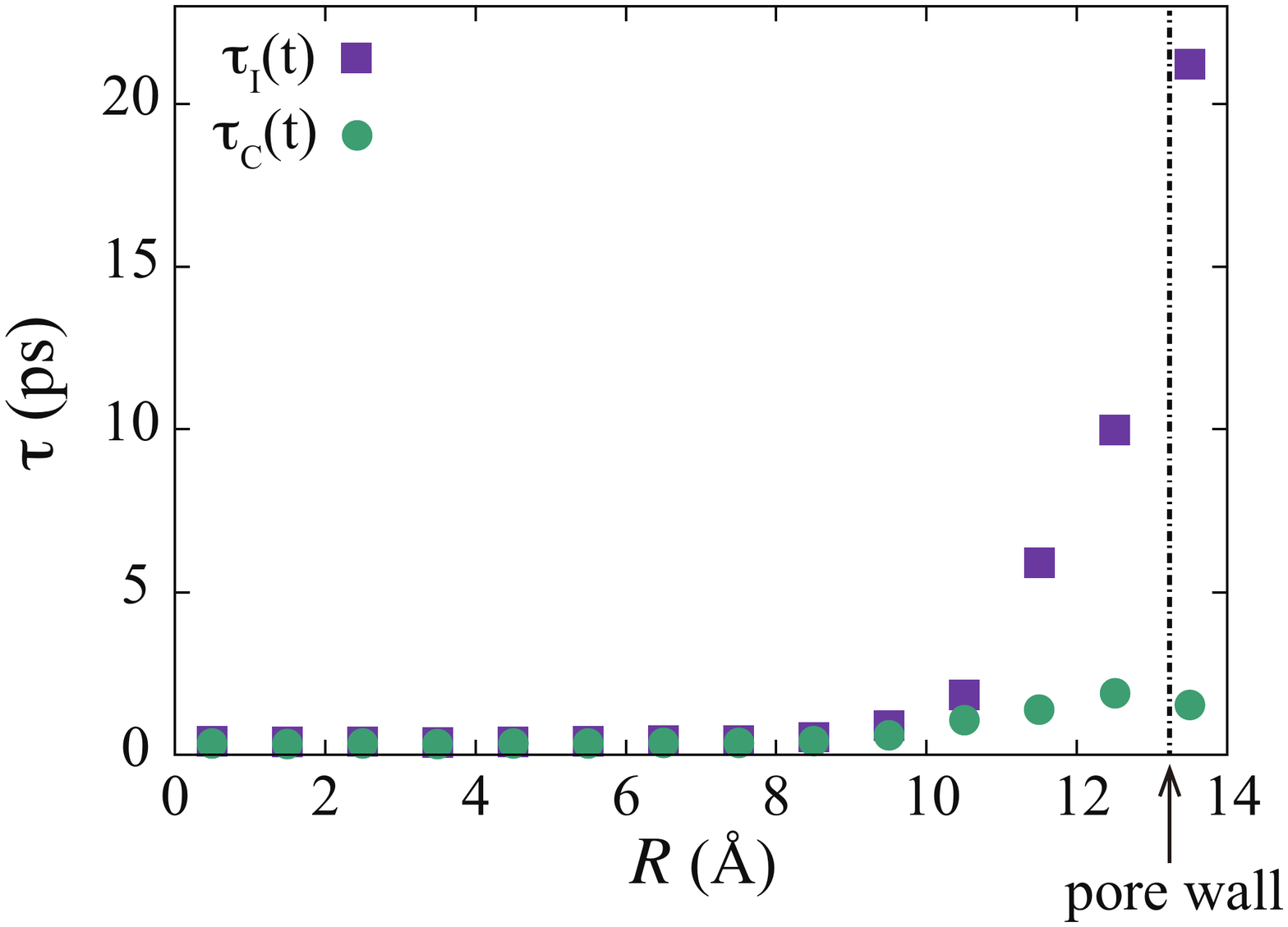}
\caption{Intermittent ($\tau_{\rm I}$) and continuous ($\tau_{\rm C}$) H-bond lifetimes shown as a function of radial distance from the pore central axis ($R$). $\tau_{\rm I}$ and $\tau_{\rm C}$ were calculated from Eq.~\ref{eq5}. The location of the pore wall ($R_{\rm Si-in}$) is also displayed as a black dash-dotted line. }
\label{f7}
\end{figure}

\begin{figure}
\includegraphics[width=8.5cm]{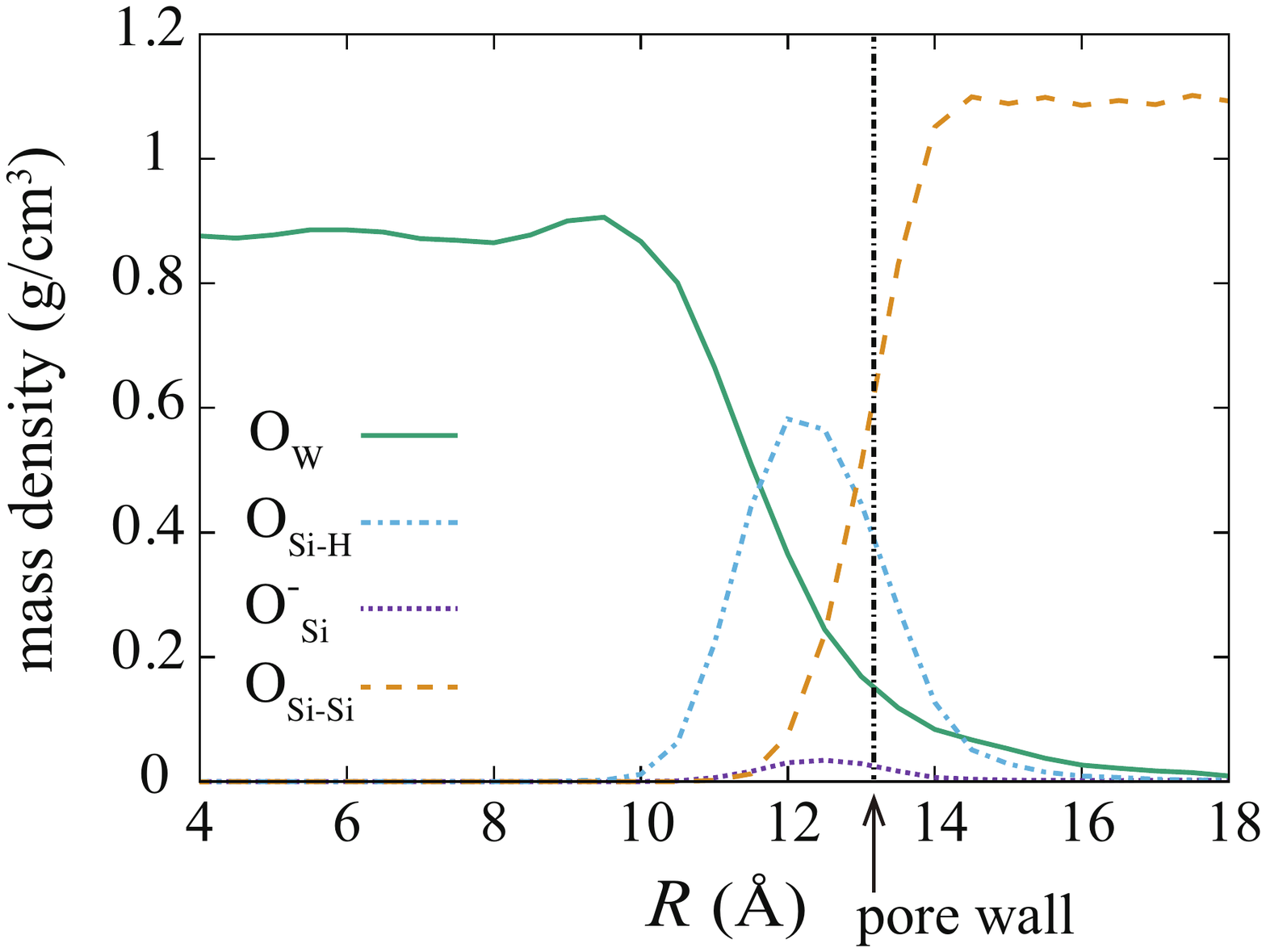}
\caption{Mass density profiles of several types of O atoms as a function of radial distance from the pore central axis ({\it R}). O$_{\rm{w}}$, O$_{\rm{Si-H}}$, O$^{-}_{\rm{Si}}$ and O$_{\rm{Si-Si}}$ represent the oxygen in H$_{\rm{2}}$O, --Si--O--H, -Si--O$^{-}$, and --Si--O--Si--, respectively. Other types of oxygen, such as OH$^{-}$ or H$_{\rm{3}}$O$^{+}$, have much smaller densities and thus are not shown.}
\label{f8}
\end{figure}

The H-bond lifetimes $\tau_{\rm I}$ and $\tau_{\rm C}$ were calculated through Eq.~\ref{eq5} using these fitting functions and are displayed in Fig.~\ref{f7}.
They are almost constant at $R <$8 \AA, but increase at $R >$8 \AA, which corresponds to the trend in $D\rm_z$ (Fig.~\ref{f4}). 
The $\tau_{\rm I}$ in the center of the pore is $\sim$0.4 ps, which is shorter than those obtained in other studies~\cite{Antipova2013,Tamai1996,Guardia2005,Padro2002}.
The $\tau_{\rm I}$ in bulk water is reported to be 2 to 9 ps depending on the water model.
One of the reasons is that $\tau_{\rm I}$ was calculated only for atoms that remain in each thin layer.
Atoms that moved to the adjacent layer were not counted in the $C_x$({\it t}) calculation if they reformed the same H-bond.
In fact, the $\tau_{\rm I}$ calculated over atoms located at $R <$6 \AA\ was about 3.2 ps, which is comparable with literature values~\cite{Antipova2013,Tamai1996,Guardia2005,Padro2002}.
Another reason is the short simulation time.
H-bonds that reformed over the 20-ps simulation were not included in the $\tau_{\rm I}$ calculation, although they would raise $\tau_{\rm I}$.
From the $\tau_x$ calculated in this work, however, the dynamical behavior of confined water can be captured.
Figure~\ref{f5} shows that $\langle n_{\rm HB} \rangle$ is almost constant up to $R$ = 11 \AA, although the H-bond dynamics begin to change from $R \sim$ 8 \AA\ (Fig.~\ref{f7}).
This is because a few H-bond pairs change from water to silanol.
The density profiles of four types of oxygen, H$_2$O, --Si--O--H, --Si--O$^{-}$, and --Si--O--Si--, are displayed as a function of $R$ in Fig.~\ref{f8}.
Other types of oxygen, such as OH$^{-}$ or H$_3$O$^{+}$, have much smaller densities and thus are not shown in the figure.
The density of silanol oxygen (O$_{\rm Si-H}$) starts to increase from $R \sim$ 10 \AA\ and peaks at $R \sim$ 12 \AA.
Thus, oxygen in water (O$_{\rm w}$) within 8 \AA$< R <$10 \AA\ can form H-bonds with silanols because the H-bond distance is 1.35--2.45 \AA.
Oxygen atoms within 8 \AA$< R <$10 \AA, where H-bonds include silanol, have an average of 3.5 H-bonds (Fig.~\ref{f4}) like those near the center of the pore, where H-bonds are formed only by water.
The decrease in the $D\rm_z$ of oxygen with $R$ at $R >$8 \AA\ (Fig.~\ref{f4}) implies that the H-bond network including silanol slows down water dynamics.
The influence of interfacial silanol on the H-bond lifetime was also reported for confined glycerol in a silica nanopore~\cite{Busselez2009}.

    $\tau_{\rm C}$ slightly increases from $R\sim$ 8 \AA\ and peaks at around $R$ = 12--13 \AA.
To see the first breakage time of H-bond more clearly, we calculated another time constant, $\tau_{\rm{on}}$.
It can be evaluated from $P_{\rm{on}}(t)$~\cite{Tamai1996}, which is defined as the distribution of time from the formation of a H-bond to its destruction. $\tau_{\rm{on}}$ is calculated using the following equation,
\begin{equation}\label{eq7}
\tau_{\rm on}=\int_0^{\infty} tP_{\rm on}(t)dt.
\end{equation}

\begin{figure}
\includegraphics[width=8cm]{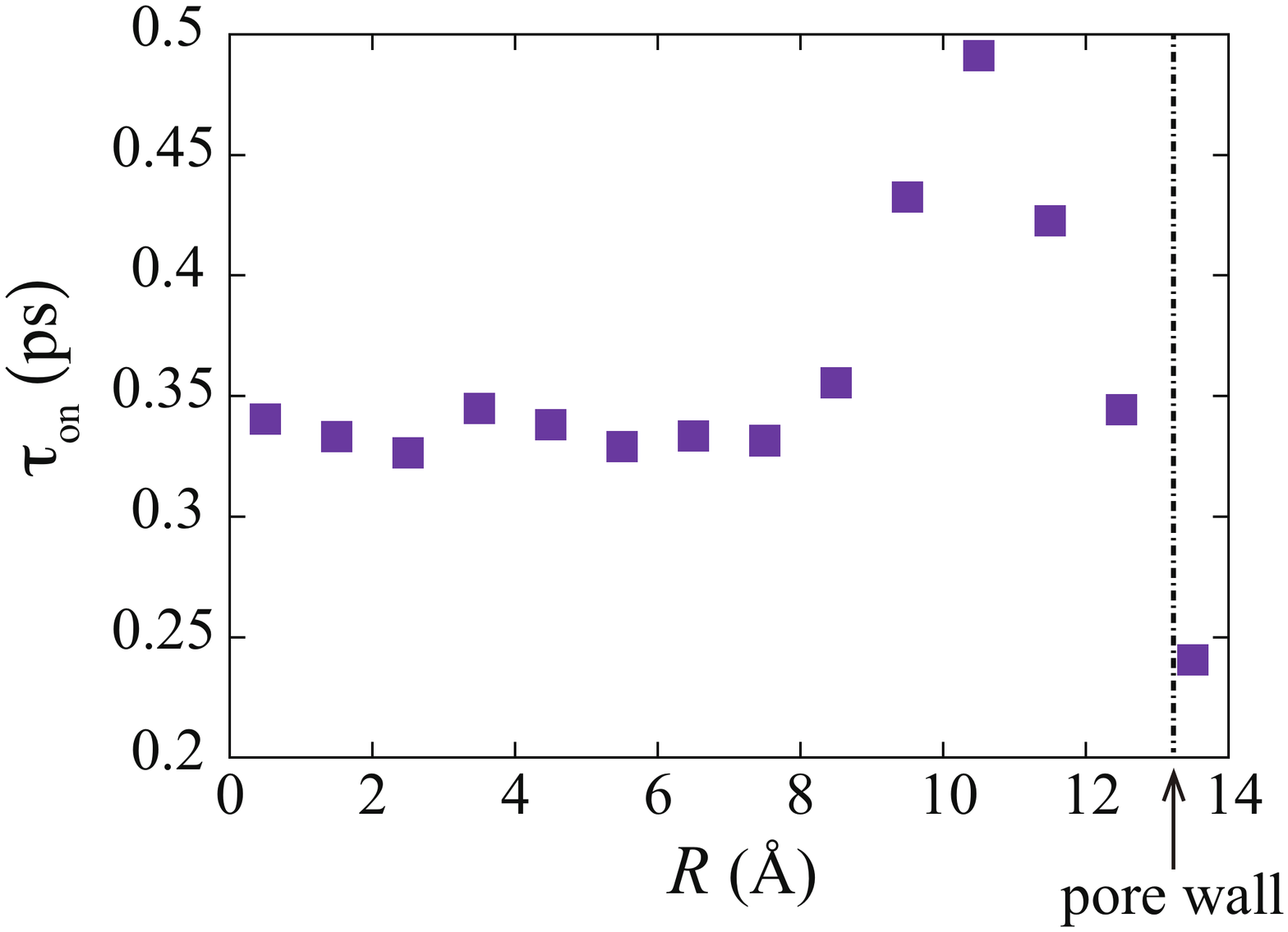}
\caption{Average hydrogen bond lifetime ($\tau_{\rm on}$) as a function of radial distance from the pore central axis ({\it R}). $\tau_{\rm on}$ was calculated using Eq.~\ref{eq7}. The location of the pore wall ($R_{\rm Si-in}$) is also displayed as a black dash-dotted line.}
\label{f9}
\end{figure}

Figure~\ref{f9} shows $\tau_{\rm on}$, which was calculated over a 50-ps time period.
At $R <$8 \AA, $\tau_{\rm on}$ is almost constant at $\sim$0.34 ps, which is comparable to that of bulk water in Ref.~\cite{Tamai1996}. 
$\tau_{\rm{on}}$ also increases with $R$ in the range of 8 \AA$< R <$11 \AA, but decreases with $R$ at $R >$11 \AA.
This behavior is similar to that of $\tau_{\rm C}$, although the peak position is slightly different.
The behavior of $\tau_{\rm on}$ at $R >$8 \AA\ is elucidated by considering the valence unit per hydrogen bond ({\it vu}), which was introduced as an index of H-bond strength by Machesky {\it et al.}~\cite{Machesky2008}.

$vu$  is obtained from
\begin{equation}\label{eq8}
vu = 1.55 - 1.06 r_{\rm avg} + 0.186 r_{\rm avg}^2,
\end{equation}
where the average H-bond distance $r_{\rm avg}$ is calculated as
\begin{equation}\label{eq9}
r_{avg} = \frac{\int_{r=1.35}^{2.45} r g_{\rm OH}(r) r^2 dr}{\int_{r=1.35}^{2.45} g_{\rm OH}(r) r^2 dr}.
\end{equation}

\begin{figure}
\includegraphics[width=14cm]{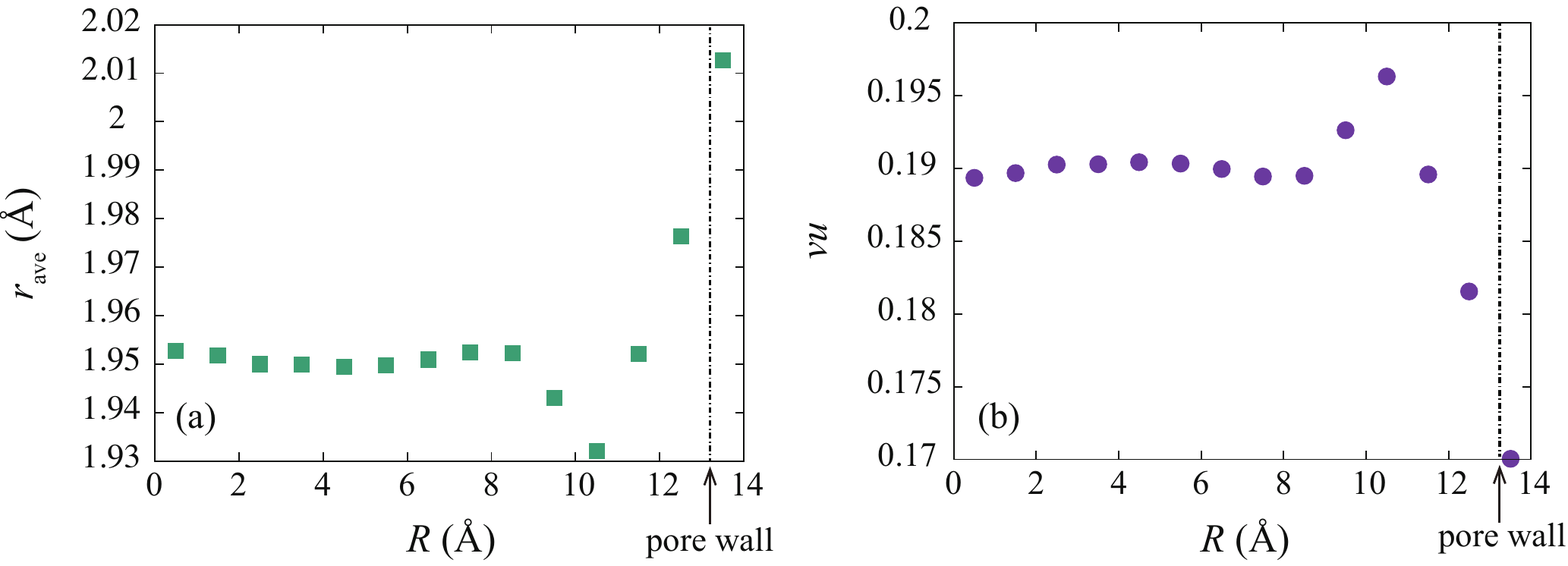}
\caption{(a) Average H-bond distance ($r_{\rm{avg}}$) and (b) valence units per hydrogen bond ($vu$) as a function of radial distance from the pore central axis ($R$). The location of the pore wall ($R_{\rm Si-in}$) is also displayed as a black dash-dotted line. }
\label{f10}
\end{figure}

$g_{\rm OH}(r)$ in Eq.~\ref{eq9} is calculated for oxygen atoms in each layer with a width of 1.0 \AA.
The integration range of $r$ = 1.35--2.45 \AA\ corresponds to the second peak of $g_{\rm OH}$, namely, the nearest intermolecular O--H distance.
Figure~\ref{f10} shows $r_{\rm avg}$ and $vu$ as a function of $R$.
$vu$ at $R <$8 \AA\ is about 0.19, which agrees with the value for a normal H-bond in bulk water at 298 K (0.2)~\cite{Brown2002}.
The plots in Figs.~\ref{f9} and~\ref{f10} suggest that the behavior of $vu$ against $R$ is similar to that of $\tau_{\rm on}$.
$vu$ increases between 8 and 11 \AA\ and reflects the decrease in $r_{avg}$ (Fig.~\ref{f10}), which indicates that H-bonds including silanol (8 \AA$< R <$11 \AA) are stronger than that of bulk water.
Stronger H-bonds lead to a longer H-bond lifetime (Fig.~\ref{f9}) and slower diffusion (Fig.~\ref{f4}); thus, H-bonds involving silanol are long lived.

On the other hand, $D\rm_z$ values at $R >$11 \AA\ are quite small compared with those at $R <$11 \AA\ (Fig.~\ref{f4}), which indicates that molecules located at $R >$11 \AA\ are almost immobile.
Moreover, at $R >$11 \AA, $\tau_{\rm{I}}$ (i.e., the final breakage time) increases with $R$ (Fig.~\ref{f7}), while $\tau_{\rm on}$ (i.e., the first breakage time) decreases with $R$ (Fig.~\ref{f9}).
These observations suggest that the H-bond acceptor (O) at $R >$11 \AA\ frequently and repeatedly forms and destroys the H-bond with the same H-bond donor (OH).
This is caused by the less-structured H-bond network, as shown in Fig.~\ref{f5}.
Fig.~\ref{f2}(a) shows that the density distribution of Si starts to increase at $R =$ 11 \AA\ and continues until $R =$ 14 \AA.
The density distribution of O in SiO$_2$ also increases within this region, as shown in Fig.~\ref{f8}.
Namely, the roughness of a-SiO$_2$ surface disrupts the formation of H-bond network.

\begin{figure}
\includegraphics[width=12cm]{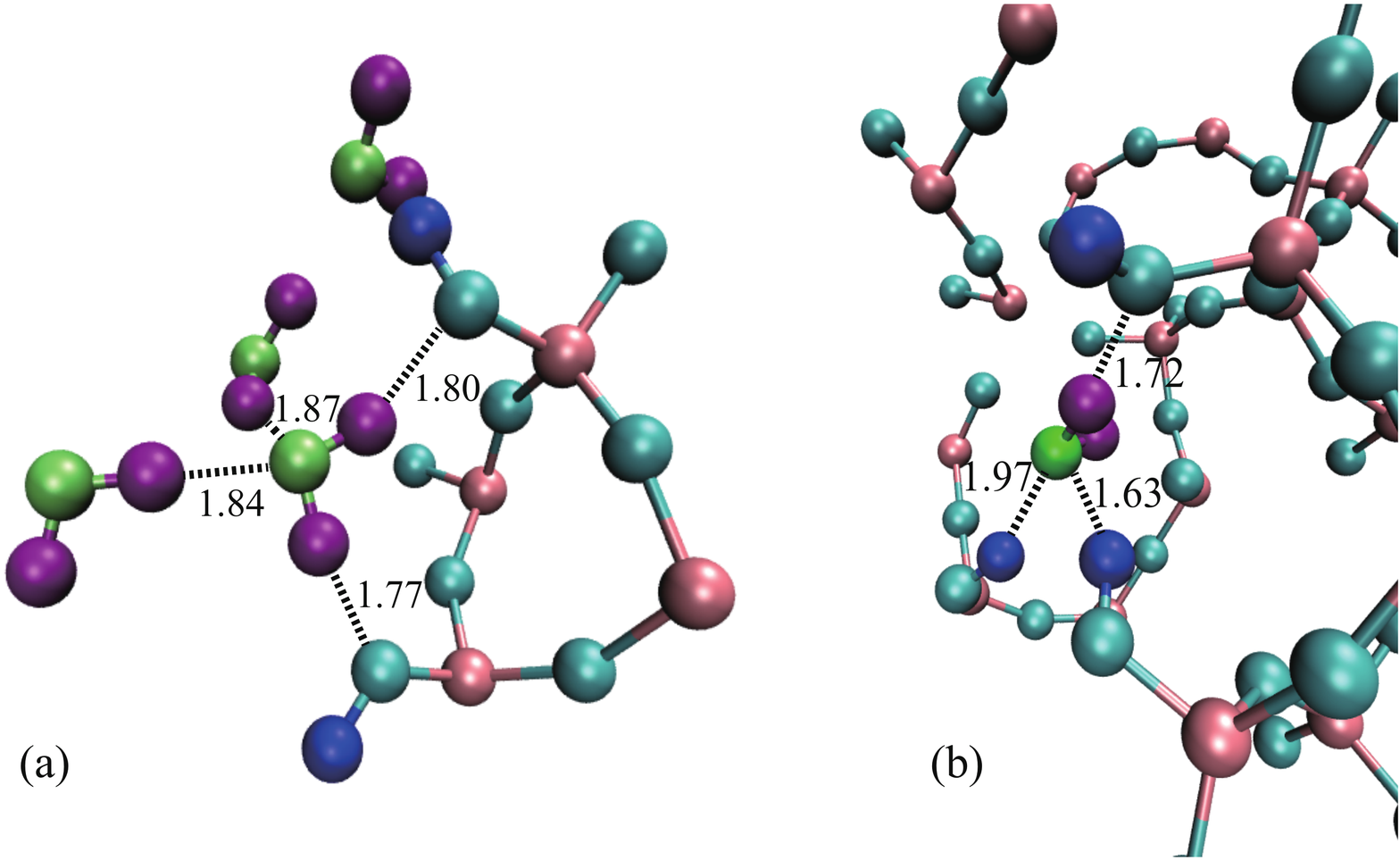}
\caption{ H-bond network of interfacial water located at $R =$ (a) 10 and (b) 13 \AA~[silicon (pink), silica oxygen (cyan), silanol hydrogen (blue), water oxygen (green), water hydrogen (purple)]. Hydrogen bonds are shown with thick dashed lines, and values next to the lines indicate the H-bond lengths (\AA).}
\label{f11}
\end{figure}

\begin{figure}
\includegraphics[width=8cm]{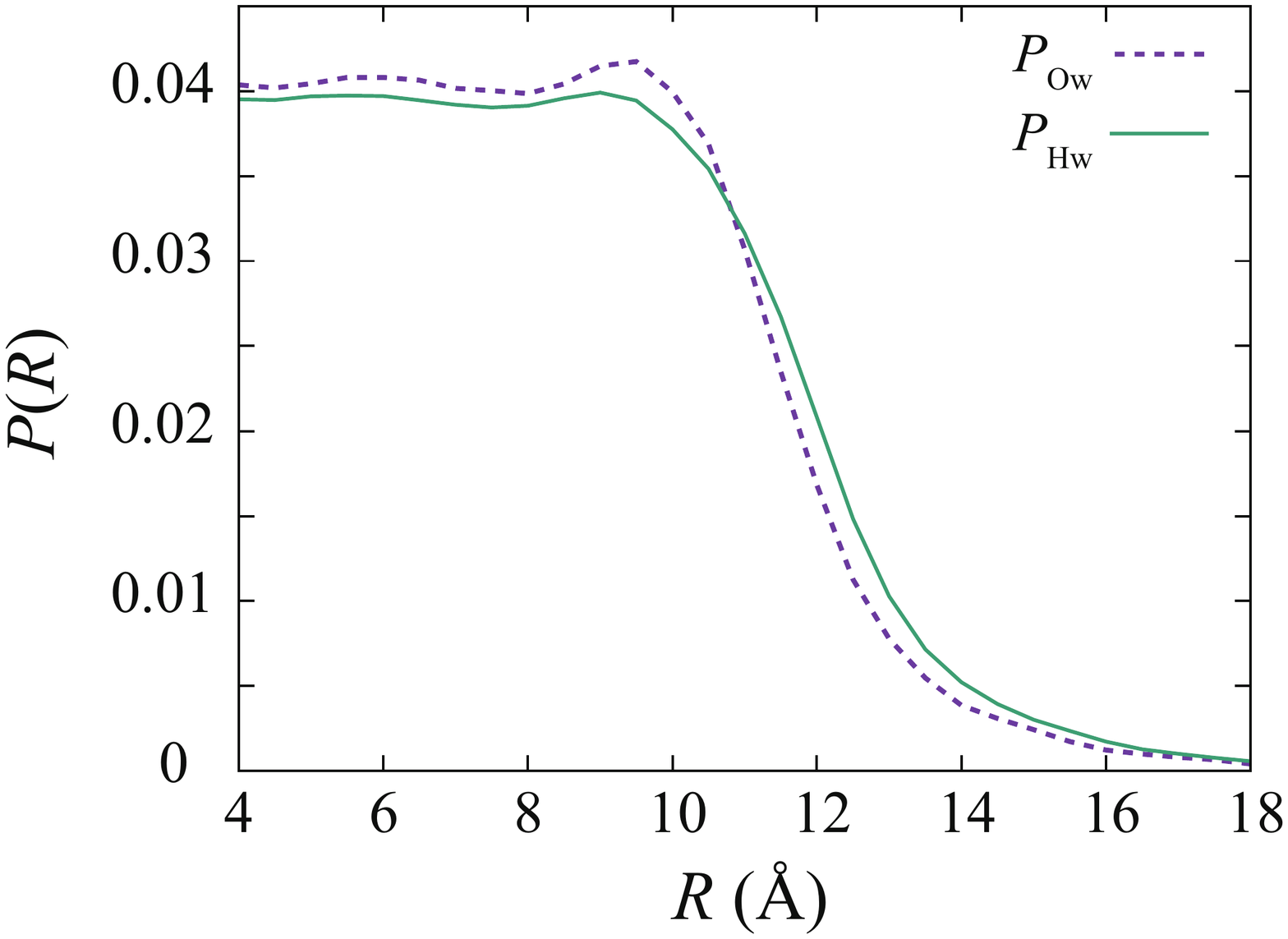}
\caption{The distribution of water oxygen $P_{\rm Ow}$ (dashed line) and water hydrogen $P_{\rm Hw}$ (solid line). }
\label{f12}
\end{figure}

Finally, we focused on the H-bond network and surface topology of water located at $R =$ 10 \AA\ (Fig.~\ref{f11}(a)) and $R =$ 13 \AA\ (Fig.~\ref{f11}(b)).
Water at $R =$ 10 \AA\ forms four H-bond networks, two each with water and silanol.
The H-bond lengths with silanol are 1.77 and 1.80 \AA, while those with water are 1.84 and 1.87 \AA.
The shorter (and stronger) H-bond length with silanol attracts water near the interface and leads to a slight increase in the density of O$_{\rm w}$ around $R =$ 8--10.5 \AA\ (Fig.~\ref{f8}), which also contributes to the slower diffusion.
Water molecules at $R =$ 13 \AA\, on the other hand, are located in a small hole on the rough silica surface and surrounded by three silanols.
It is separated from the water H-bond network and form H-bonds only with silanols.
The configuration of water at $R>$11 \AA\ can be estimated from Fig.~\ref{f12}.
It shows the distribution of water atoms $P_{\rm Ow}$ and $P_{\rm Hw}$, normalized so that the sum of each probability becomes 1.
At $R>$11 \AA, $P_{\rm Hw}$ is slightly larger than $P_{\rm Ow}$.
It indicates that water hydrogen is closer to the wall than water oxygen in this region.
Water configuration at $R>$11 \AA\ is thus presumed to be captured in a silica hall and with silanols as shown in Fig.~\ref{f11}(b).
Because of this strong H-bond network with silanols and steric hindrance by surrounding silica atoms, they are almost immobile.

\section{\label{sec:discuss}Summary and Conclusion}
We conducted large-scale MD simulations of confined water in nanoporous silica with a pore diameter of 2.7 nm using ReaxFF.
We compared Yeon's and Pitman's parameter sets for our system.
We found that the former is suitable for the analyses of water structure and dynamics in a 2.7-nm silicate pore, although the density of a-SiO$_2$ was better using Pitman's parameter set.
We calculated the radial distribution function and diffusion coefficient of water using Yeon's parameter set, and the values in the center of the pore agreed well with those of real water obtained from several experimental studies.
In addition, the diffusion coefficient of water essentially corresponded to previous simulation studies using the classical MD force fields, CLAYFF and SPC/E water~\cite{Bourg2012}.
Thus, these force fields can capture the water diffusion in silica nanopores well.

We divided the pore into hollow cylindrical layers and calculated the diffusion coefficient, average number of H-bonds, H-bond lifetime, and H-bond strength as a function of the radial distance from the pore central axis ($R$).
Based on H-bond structure and dynamics, the pore can be divided into three regions: $R <$8 \AA, 8 \AA$< R <$11 \AA, and $R >$11 \AA.
At $R <$8 \AA, water behaves like bulk water.
Within 8 \AA$< R <$11 \AA, H-bonds involving silanol have a longer H-bond lifetime, although the average number of H-bonds per atom does not change from that in the pore center.
H-bonds involving silanol become stronger, resulting in a longer H-bond lifetime and smaller  diffusion coefficient.
Finally, at $R >$11 \AA, the H-bond network becomes less structured owing to fewer surrounding water.
This is caused by steric hindrance by interfacial Si and O in SiO$_2$ and surrounding silanols, which arise from the amorphous surface.
Thus, the elapsed time for the final H-bond breakage ($\tau_{\rm I}$) becomes longer and the diffusion coefficient becomes smaller.

Water behavior near the interface depends on the roughness of the surface and fraction of silanols present, which will be investigated in our future work.
The MD simulation with ReaxFF provides highly reliable results, as evidenced by the good agreement with experimental values.
Thus, this method is also expected to accurately predict water behavior near the interface, which is difficult to access in experiments.
In addition, H-bonds involving silanol become stronger than that of bulk water because of the shorter O--H distance, although the detailed structure of the H-bond network is unclear.
In future work, we will also examine how the network structure causes slower dynamics in detail.

\section*{Acknowledgement(s)}
This research used the computational resources of the K computer provided by the RIKEN Advanced Institute for Computational Science through the HPCI System Research project (Project ID: hp170021), of the Supercomputer Center of the Institute for Solid State Physics provided by The University of Tokyo, and of the Research Center for Computational Science at Okazaki, Japan.
The authors are partially supported by the Research Center for Computational Science, National Institute of Natural Sciences. 
This work was partially supported by the Japan Society for the Promotion of Science KAKENHI grant nos. JP15K05244, JP18H04494 and JP19K05209.

\end{document}